%
\magnification 1200
\hsize 145 true mm
\vsize 226 true mm
\magnification=\magstep1
\baselineskip=18truept
\parindent=20truept
\parskip=0truept
\rightline{MIT-CTP-2606} 
\medskip
\centerline{\bf Nonlocal Condensates in QCD}
\medskip
\centerline{W-Y. P. Hwang}
\smallskip
\parindent=0truept
\centerline{Department of Physics, National Taiwan University}

\centerline{Taipei, Taiwan 10764, R.O.C.}

\centerline{and}

\centerline{Center for Theoretical Physics}

\centerline{Laboratory for Nuclear Science and Department of Physics}

\centerline{Massachusetts Institute of Technology}

\centerline{Cambridge, Massachusetts 02139}
\smallskip
 
\bigskip
\centerline{\bf Abstract}
\medskip
\parindent=20truept
\baselineskip=24truept

In the presence of the nontrivial QCD ground state or vacuum, nonlocal 
condensates are used to characterize the quark or gluon propagator, or 
other Green functions of higher order. In this paper, we wish to show 
that, by taking the large $N_c$ limit (with $N_c$ the number of color) 
in treating higher-order condensates, a closed set of coupled 
differential equations
may be derived for nonlocal condensates. As a specific example,
the leading-order equations for the nonlocal condensates appearing in
the quark propagator are derived and explicit solutions are obtained.  
Some applications of our analytical results are briefly discussed.
\vskip 2cm
\parindent=0truept
PACS Indices: 12.38.Lg; 12.38.Cy; 12.38.Aw.
 
\vfill\eject
 
\parindent=20truept
\centerline{\bf I. Introduction}
\medskip

The problem of strong interaction physics has been around for more than 
half a century, but the very nature of the problem varies 
with the so-called ``underlying theory'' which nowadays is taken
universally to be quantum chromodynamics or QCD. Although the asymptotically 
free nature of QCD allows us to test the candidate theory at high energies, 
the nonperturbative feature dominates for hadrons or nuclei at low 
energies. As of today, it remains 
almost impossible to solve problems related to hadrons or nuclei, except 
perhaps through lattice simulations but it is unlikely that, through the 
present-day computation power which is still not quite adequate, 
simulations would give us most of what we wish to know. However, the
need of solving QCD directly has gained tremendous importance over
the last decade and, to this end, any fruitful attempt in confronting 
directly with QCD should be given due attention.  

The ground state, or the vacuum, of QCD is known to be nontrivial, in the 
sense that there are non-zero condensates, including gluon condensates, 
quark condensates, and perhaps infinitely many higher-order condensates. 
In such a theory, propagators, i.e. causal Green's functions, such as the 
quark propagator
$$ iS_{ij}^{ab}(x) \equiv <0\mid T(q_i^a(x) {\bar q}_j^b(0))\mid 0>, 
\eqno(1)$$
carry all the difficulties inherent in the theory. Higher-order condensates,
such as a four-quark condensate,
$$ <0\mid T({\bar\psi}(z) \gamma_\mu \psi (z) q_i^a(x) {\bar q}_j^b(0))\mid 0>,
$$
with $\psi(z)$ also labeling a quark field, represent an infinite series of
unknowns unless some useful ways for reduction can be obtained. As the vacuum,
$\mid 0>$, is highly nontrivial, there is little reason to expect that Wick's
theorem (of factorization), as obtained for free quantum field theories, is 
still of validity. Thus, we must look for alternative methods in order to 
obtain useful results.

It is known that the equations for Green functions up to a certain order 
usually involve Green functions of even higher order, thereby making such 
hierarchy of equations often useless in practice. In this paper, however, 
we wish to show that, provided that we may use the large $N_c$ 
approximation to treat condensates of much higher order, there 
is in fact a natural way of setting up closed sets of differential 
equations which govern the inter-related Green functions to a given order. 
We consider this as an important accomplishment, both because we can
always go over to the next level of sophistication in order to improve
the approximation and because the large $N_c$ expansion has been shown to
yield desirable results for describing hadron physics.  

\bigskip
\bigskip
\centerline{\bf II. Leading-order Equations for Nonlocal Condensates}
\medskip
\medskip

In light of the nontrivial QCD vacuum, we begin by considering the 
feasibility of working directly with the various matrix elements such 
as the quark propagator of Eq. (1). Useful relations may be derived 
if we regard the equations for interacting fields [1],
$$\eqalignno{
&\{i \gamma^\mu (\partial_\mu + ig {\lambda^a\over 2} A_\mu^a) - m \}\psi =0; 
 &(2)\cr
&\partial^\nu G_{\mu\nu}^a - 2 g f^{abc} G^b_{\mu\nu}A^c_\nu 
 +g {\bar \psi} {\lambda^a\over 2}\gamma_\mu \psi = 0, &(3)}$$
as the equations of motion for {\it quantized} interacting fields,
subject to the standard rule for quantization that the equal-time 
(anti-)commutators among these quantized interating fields are identical to
those among non-interacting quantized fields. As our basic example, 
we allow the operator $\{i \gamma^\mu \partial_\mu - m \}$ to act on the 
matrix element defined by Eq. (1) and obtain
$$\eqalignno{
\{i \gamma^\mu \partial_\mu - m \}_{ik} iS_{kj}^{ab} (x) 
  = & i \delta^4(x) \delta^{ab} \delta_{ij} 
     +<0\mid T(\{g {\lambda^n\over 2}A_\mu^n\gamma^\mu q(x)\}_i^a 
     {\bar q}_j^b(0))\mid 0>. &\cr &&(4)}$$
We should always keep in mind that the QCD vacuum $\mid 0>$ is a 
nontrivial ground state which is in general not annihilated by operating
on it the annihilation operators. 
  
Eq. (4) can be solved by splitting the propagator into a singular,
perturbative part and a nonperturbative part:
$$iS_{ij}^{ab} (x) = i S_{ij}^{(0)ab}(x) + i {\tilde S}_{ij}^{ab}(x), 
\eqno(5)$$
where
$$\eqalignno{
  iS_{ij}^{(0)ab}(x) &\equiv \int {d^4p\over (2\pi)^4} e^{-ip\cdot x}   
    iS_{ij}^{(0)ab}(p), &(6a)\cr
  iS_{ij}^{(0)ab}(p) &= \delta^{ab} {i ({\hat p} + m)_{ij}\over p^2 -m^2 +i
   \epsilon}, &(6b)}$$
with ${\hat a} \equiv \gamma^\mu a_\mu$ for a four-vector $a_\mu$.
The nonperturbative part then satisfies the equation:
$$\{i \gamma^\mu \partial_\mu - m \}_{ik} i{\tilde S}_{kj}^{ab} (x) 
  =  <0\mid T(\{g {\lambda^n\over 2}A_\mu^n\gamma^\mu q(x)\}_i^a 
     {\bar q}_j^b(0))\mid 0>. \eqno(7)$$
This would be pretty much the end of the story unless we could find some
way to proceed. We should note that Eq. (4) may also be derived by 
making use of, e.g., the path-integral formulation, and the issue of
how to define renormalized composite operators, i.e. products of field
operators, is by no means trivial (and fortunately we need not worry
about such problem for the sake of this paper).  

As a useful benchmark, we note that, with the fixed-point gauge, 
$$ A_\mu^n(x) = -{1\over 2} G_{\mu\nu}^n x^\nu + \cdot\cdot\cdot, 
\eqno(8)$$
we may solve the nonperturbative part $i{\tilde S}_{ij}^{ab}(x)$ as a
power series in $x^\mu$, 
$$\eqalignno{
i{\tilde S}_{ij}^{ab}(x) = & -{1\over 12} \delta^{ab}\delta_{ij} <{\bar q}q>
   + {i\over 48} m {\hat x}_{ij}\delta^{ab} <{\bar q}q> &\cr
    &+{1\over 192} <{\bar q} g \sigma\cdot G q> \delta^{ab} x^2 \delta_{ij}
     +\cdot\cdot\cdot, &(9)}$$
The first term is the integration constant which defines the so-called
``quark condensate'', while the mixed quark-gluon condensate appearing in
the third term arises because of Eqs. (7) and (8). It is obvious that 
the series (9) is a short-distance expansion, which converges
for small enough $x_\mu$. We note that is just the standard 
quark propagator cited in most papers in QCD sum rules [2]. 

The approach which we suggest here [3] is based upon two key elements, 
namely, the set of interacting field equations {\it plus} the rule of 
canonical quantization (for interacting fields). The equations which we 
obtain, such as Eq. (5), are much the same as
the set of Schwinger-Dyson equations (for the matrix elements). An important
aspect in our derivation is that the nontriviality of the vacuum $\mid 0>$ is
observed at every step $-$ a central issue in relation to QCD. 

As another important exercise, we may split the gluon propagator into the 
singular, perturbative part and the nonperturbative part. The 
nonperturbative part is given by [3]
$$\eqalignno{
 & g^2 <0\mid \colon G_{\mu\nu}^n(x) G_{\alpha\beta}^m(0) \colon \mid 0> &\cr
= &\quad {\delta^{nm}\over 96} <g^2G^2> (g_{\mu\alpha}g_{\nu\beta} 
    -g_{\mu\beta}g_{\nu\alpha}) &\cr
  &- {\delta^{nm}\over 192} <g^3 G^3> \{ x^2 (g_{\mu\alpha}g_{\nu\beta} -
  g_{\mu\beta}g_{\nu\alpha}) - g_{\mu\alpha}x_\nu x_\beta + g_{\mu\beta}x_\nu
    x_\alpha -g_{\nu\beta}x_\mu x_\alpha +g_{\nu\alpha} x_\mu x_\beta \} &\cr
  & + O(x^4), &(10)}$$
with 
$$\eqalignno{
 <g^2 G^2 > & \equiv <0\mid \colon g^2 G_{\mu\nu}^n (0) G^{n\mu\nu}(0)
    \colon \mid 0>, & (11a)\cr
 <g^3 G^3> & \equiv <0\mid \colon g^3 f^{abc}g^{\mu\nu} G_{\mu\alpha}^a(0)
    G^{b\alpha\beta}(0) G_{\beta\nu}^c(0) \colon \mid 0>.&(11b)}$$
Again, the first term in Eq. (10) is the integration constant for the 
differential equation satisfied by the gluon propagator. Note that inclusion
of the gluon condensate $<g^2 G^2 >$ in Eq. (10) is standard but 
the triple gluon condensate $<g^3 G^3 >$ is a new entry required by the 
interacting field equation (3). Our approach indicates when condensates of 
entirely new types should be introduced as we try to perform operator-product 
expansions to higher dimensions.

In what follows, we wish to focus on the quark propagator, as specified 
by Eqs. (1) and (4)-(7). We write
$$ i{\tilde S}_{ij}^{ab}(x) = \delta^{ab}\{ \delta_{ij} f(x^2) + 
i{\hat x}_{ij} g(x^2) \}, \eqno(12)$$
with $x^2 \equiv x_0^2 - \rm{\bf x}^2$. $f(x^2)$ and $g(x^2)$ are what 
we refer to as ``nonlocal condensates'' in connection with the quark
propagator. We note that 
$$<:{\bar q}(x)\,q(0):> = -12 f(x^2),\eqno(13)$$
which is the nonlocal quark condensate in the standard sense. To proceed
further, we work only with the leading term in the fixed-point gauge and 
introduce
$$\eqalignno{
&<:\{g G_{\mu\nu} q(x)\}^a_i {\bar q}^b_j(0):>&\cr
=&\delta^{ab} \{ (\gamma_\mu x_\nu - \gamma_\nu x_\mu) A(x^2)
                +i \sigma_{\mu\nu} B(x^2) 
       + (\gamma_\mu x_\nu - \gamma_\nu x_\mu){\hat x} C(x^2)
                +i \sigma_{\mu\nu} {\hat x} D(x^2) \}, &(14)}$$
with $G_{\mu\nu} \equiv (\lambda^n/2)G^n_{\mu\nu}$, an antisymmetric 
operator. The invariant 
functions $A(x^2)$, $B(x^2)$, $C(x^2)$, and $D(x^2)$ are additional
nonlocal condensates which we deal with explicitly in this paper. 

Under the assumption that we keep only the leading term in the fixed-point
gauge (a simplifying assumption which can be removed whenever 
necessary), we have
$$\eqalignno{
 &  \{i \gamma^\alpha \partial_\alpha -m\}_{ik}
   <:\{ g G_{\mu\nu} q(x)\}^a_k {\bar q}^b_j(0):>&\cr
=& -{1\over 2} x^\beta <:\{g^2 G_{\mu\nu} G_{\alpha\beta} \gamma^\alpha
   q(x)\}^a_i {\bar q}^b_j(0):> &(15a)\cr
=& -{1\over 2}\cdot {1\over 96}\cdot {4\over 3}\cdot <g^2 G^2>
   <:\{(\gamma_\mu x_\nu - \gamma_\nu x_\mu)q(x)\}^a_i {\bar q}^b_j(0):>
   &(15b)\cr
=& -{1\over 144} <g^2 G^2> \delta^{ab} (\gamma_\mu x_\nu -\gamma_\nu
       x_\mu) \{ f(x^2) + i {\hat x} g(x^2) \}. &(15c)}$$
Here the second line, (15a), follows from the field equation and the 
third line, (15b), is based on the large $N_c$ approximation that 
the contribution in which $G_{\mu\nu}$ and $G_{\alpha\beta}$ do not 
couple to color-singlet is suppressed by a factor of $1/N_c^2$. Thus,
the factorization in the present case is justified to order $1/N_c^2$, 
rather than just order $1/N_c$. We also note that we have used Eq. (10)
to leading order in obtaining Eq. (15b), but this approximation can be
relaxed if necessary.

Now, we may use Eqs. (7) and (15c) and obtain a closed set of equations:
$$\eqalignno{
& 2 f'(x^2) - m g(x^2) = -{3\over 2} i (B - x^2 C), &(16a)\cr
& 2 x^2 g'(x^2) + 4 g(x^2) + m f(x^2) = {3\over 2} x^2 (A-D), &(16b)\cr
& 4 iB' - 2iC - 2ix^2 C' - mA = - {1\over 144} <g^2 G^2> f(x^2),
  &(16c)\cr
& 2iA +2i x^2 D' - mB =0, &(16d)\cr
& - 2iA' +4iD' -mC = - {i\over 144} <g^2 G^2> g(x^2), &(16e)\cr
& 2i B' +2iC -mD =0, & (16f)}$$
where the derivatives are with respect to the variable $x^2$.

Treating $m$ as an expansion parameter, 
$$F(x^2) = \sum_{k=0}^\infty m^k F_k(x^2), \eqno(17)$$
we may solve the coupled equations, (16a)-(16f), order by order in $m$.
To leading order in $m$, we obtain
$$\eqalignno{
&x^2 f_0^{\prime\prime\prime}+3 f_0^{\prime\prime} - \xi_0^2 x^2 f'_0 
  - 2 \xi_0^2 f_0 = 0, &(18)\cr
&(x^2)^3 g_0^{\prime\prime\prime} + 5 (x^2)^2 g_0^{\prime\prime} 
   +\{2x^2 - \xi_0^2 (x^2)^3\} g'_0 -\{2+ 2\xi_0^2 (x^2)^2\}g_0 =0, 
    &(19)}$$
with $\xi_0^2 \equiv <g^2 G^2>/384$. The equations for $A_0$, $B_0$,
$C_0$, and $D_0$ can easily be solved once we obtain $f_0$ and $g_0$.

Eq. (19) can be simplified considerably by introducing
$$g_0(x^2) \equiv (x^2)^{-2} {\tilde g}_0(x^2), \eqno(20a)$$
which leads to the equation:
$$ x^2 {\tilde g}_0^{\prime\prime\prime} - {\tilde g}_0^{\prime\prime}
   -\xi_0^2 x^2 {\tilde g}_0^\prime =0. \eqno (20b)$$

Eqs. (17) and (20) can be solved by iteration, leading to the result:
$$\eqalignno{
 f_0(t) =& a_0 \{ 1+ {1\over 1\cdot 3} (\xi_0 t)^2 + {1\over 1\cdot 3^2
     \cdot 5} (\xi_0 t)^4 +\cdot\cdot\cdot \} &\cr
        +& a_1 t \{ 1 +{1\over 2\cdot 4} (\xi_0 t)^2 + {1\over 2\cdot 4^2
     \cdot 6} (\xi_0 t)^4 +\cdot\cdot\cdot \}, &(21)\cr
 {\tilde g}_0^\prime (t) =& c_2 t^2 \{ 1 + {1\over 2\cdot 4} (\xi_0 t)^2
     +{1\over 2\cdot 4^2\cdot 6} (\xi_0 t)^4 + \cdot\cdot\cdot \},
     &(22)}$$
with $t \equiv x^2$ and 
$$ a_0 = -{1\over 12} <{\bar q}q>, \qquad a_1 = {1\over 192} <{\bar q}
    g_c \sigma \cdot G q>, \qquad c_2 = -{g_c^2 <{\bar q} q>^2 
     \over 2^5 \cdot 3^4}. \eqno (23)$$
Eq. (23) is obtained by comparing to the well-known series expansion for
the quark propagator (see, e.g., [4]). Note that there are two integration
constants, $a_0$ and $a_1$, for $f(x^2)$ but there is only one permissible
constant for $g(x^2)$ (and $c_2$ is in fact a four-quark condensate taken
in the large $N_c$ limit).

For a number of applications, it is useful to obtain analytic expressions
for $f_0(t)$ and $g_0(t)$. This turns out to be possible by way of Laplace
transforms. 
$${\bar f}_0(s) \equiv \int_0^\infty ds e^{-st} f_0(t),\qquad
  {\bar {\tilde g}}_0^\prime(s) \equiv \int_0^\infty ds e^{-st} 
  {\tilde g}_0^\prime (t). \eqno(24)$$
We obtain
$$\eqalignno{
 {\bar f}_0(s) & = - {2a_1\over \xi_0^2} - {a_0\over \xi_0} {s\over
 \sqrt {s^2 -\xi_0^2}} sec^{-1} {s\over \xi_0} + {\gamma_0 s \over
 \sqrt {s^2 -\xi_0^2}}, &(25)\cr
 {\bar {\tilde g}}_0^\prime(s)& = {2c_2 \over (s^2-\xi_0^2)^{3/2}},
  &(26)}$$ 
with $\gamma_0 = 2 a_1/\xi_0^2$. Looking up the table for Laplace transforms,
we find
$${\tilde g}_0^\prime = {2c_2\over \xi_0^2}\cdot \xi_0 t\cdot I_1(\xi_0 t),
  \eqno (27)$$
with $I_1(z)$ the modified Bessel function of the first kind, to order 
one. It is straightforward to show that Eq. (27) yields the series 
expansion in Eq. (22). Also, the function $I_1(\xi_0 t)$ enters the 
second series in $f_0(t)$ as in Eq. (21).

To close our presentation of the explicit solution to leading order in
$m$, we note that Eqs. (16) yields
$$\eqalignno{
f_0^\prime (t) & = - i {3\over 4} (t B_0(t))^\prime, &(28a)\cr
C_0 (t) &= - B'_0(t), &(28b)\cr
 t g_0^\prime(t) + 2 g_0(t) & = - {3\over 4} t (tD_0(t))^\prime, 
   &(28c)\cr
 A_0(t) & = -t D_0^\prime (t).&(28d)}$$
Thus, the functions $A_0$, $B_0$, $C_0$, and $D_0$ can be solved 
explicitly once $f_0$ and $g_0$ are known. 

On the other hand, we may go beyond the leading order in $m$ and 
obtain, as example,
$$\eqalignno{
& tf_1^{\prime\prime\prime} + 3 f_1^{\prime\prime} -\xi_0^2 t f_1^\prime
  - 2 \xi_0^2 f_1 &\cr
=& {1\over 2} tg_0^{\prime\prime} +{3\over 2} g_0^\prime 
  -{3\over 8}\{ tA_0^\prime + 2A_0 +t (tD_0)^{\prime\prime} 
  +2 (tD_0)^\prime \}; &(29a)\cr
& t^3 g_1^{\prime\prime\prime} + 5t g_1^{\prime\prime} + (2t-\xi_0^2 t^3)
  g_1^\prime - (2 + 2 \xi_0^2 t^2) g_1 &\cr
=& -{1\over 2}(t^2 f_0^{\prime\prime} +t f_0^\prime -f_0) 
   +{3t^2\over 8}(-i B_0^\prime +2iC_0 +it C_0^\prime).&(29b)}$$
These equations can also be solved by Laplace transforms. In other words,
the nonlocal condensate functions $f(x^2)$ and $g(x^2)$ can be 
analytically solved order by order in $m$. 

\bigskip
\bigskip
\centerline{\bf III. Discussion and Summary}
\medskip
\medskip

Thus far, we have described how to obtain a closed set of coupled 
equations for the nonlocal condensates which are relevant in the 
description of the quark propagator. We have also shown how these 
equations can be solved explicitly. Furthermore,
some of the assumptions underlying our equations can be relaxed and 
more elaborate equations may then be obtained. Of course, some of our
results are gauge dependent as the quark propagator (1) has been analyzed
in a specific gauge (8). Nevertheless, our primary motivation for studying
the quark propagator stems from our interest in the method of QCD sum 
rules [1], which  may be regarded as the various approaches
in which one tries to exploit the roles played by the quark and gluon 
condensates for problems involving hadrons. In this regard, results 
based on our leading-order equations are often adequate and the final
answers are in general free from the potential gauge dependent problem. 

There are several approaches of QCD sum rules towards hadron physics.
As the first approach, we may consider the Belyaev-Ioffe nucleon mass
sum rules [5,4], where the short-distance expansion for the quark propagator
is needed up to a certain (high) dimension. In this context, our 
analytical results on nonlocal condensates may not be very useful if
the resultant series converges rapidly, and in general our analytical
expressions may be used to perform further analytical analysis of the
problems as a way to improve the results obtained via short-distance
expansions (in $x_\mu$).
The second approach is to consider the response of the QCD vacuum to
some external fields, such as the method of QCD sum rules in the 
presence of an external axial field $Z_\mu(x)$ [6]. In this context, 
certain induced condensates are introduced (previously as new parameters) but 
the method offers a simple extension of the first approach in 
calculating magnetic moments, coupling constants, and other quantities
by avoiding a need to treat explicitly the three-point Green's 
functions - a need which would still involve some conceptual difficulties.  
What is of great interest is that our analytical expressions for nonlocal
condensates help to determine the induced condensates previously treated
as new parameters, thereby making the external-field QCD sum rule method
more powerful than what it used to be. To illustrate the point, we 
consider the external axial field $Z_\mu$ with the interaction,
$$\delta {\cal L} (x) = g Z^\mu (x) {\bar q}(x) \gamma_\mu \gamma_5 
                        q(x). \eqno (30)$$
For a constant $Z^\mu$ field, there are two major induced condensates [6]:
$$<0\mid {\bar q}(0) \gamma_\mu \gamma_5 q(0)\mid 0>_{Z^\alpha}
\quad \rm{and} \quad
<0\mid {\bar q}(0) g_c {\tilde G}_{\mu\nu} \gamma^\nu q(0)\mid 
0>_{Z^\alpha}.$$
We now have
$$\eqalignno{
  & <0\mid {\bar q}(0)\gamma_\mu \gamma_5 q(0)\mid 0>_{Z^\alpha}&\cr
= & i \int d^4 x g Z^\alpha(x) <0\mid T({\bar q}(x) \gamma_\alpha
  \gamma_5 q(x) {\bar q}(0) \gamma_\mu \gamma_5 q(0)) \mid 0> &\cr
= & i \int d^4 x g Z^\alpha(x) \{ Tr [i S^{(0)}(-x)\gamma_\alpha \gamma_5
    iS^{(0)}(x) \gamma_\mu \gamma_5]&\cr
&\qquad\qquad\qquad +Tr[i{\tilde S}(-x) \gamma_\alpha \gamma_5 iS^{(0)}(x)
    \gamma_\mu \gamma_5] &\cr
&\qquad\qquad\qquad +Tr[iS^{(0)}(-x) \gamma_\alpha\gamma_5 i{\tilde S}(x)
    \gamma_\mu \gamma_5] &\cr
&\qquad\qquad\qquad +Tr[i{\tilde S}(-x)\gamma_\alpha\gamma_5 i{\tilde S}(x)
    \gamma_\mu \gamma_5]\}. &(31)}$$
The first term is the one-loop result which can be regularized (e.g., in
$d$ dimensions) and, as expected for a perturbative contribution, its
finite part is 
small compared to the second and third terms. The last term, which can
be treated numerically, involves products of two condensates and it
is higher than the second or third term by at least three dimensions
(and is likely of less numerical significance).

Analogously, we have 
$$\eqalignno{
 & <0\mid {\bar q}(0) g_c {\tilde G}_{\mu\nu} \gamma^\nu q(0)
   \mid 0>_{Z^\alpha} &\cr
=& i\int d^4x g Z^\alpha(x)<0\mid T({\bar q}(x)\gamma_\alpha\gamma_5 q(x)
    {\bar q}(0) g_c {\tilde G}_{\mu\nu} \gamma^\nu q(0))\mid 0>&\cr
= & ig Z^\alpha g_c {1\over 2} \epsilon_{\mu\nu\lambda\eta}\int 
   d^4x Tr\{\gamma_\alpha\gamma_5 iS^{ba}(x)\gamma^\nu 
    <:(G^{\lambda\eta}q(0))^aq^b(x):>\}.&(32)}$$
Thus, our analytical results on $A_0$, $B_0$, $C_0$, and $D_0$ [Cf. 
Eqs. (14) and (28)] may be used to evaluate Eq. (32). 

Without going into the details (which shall be presented elsewhere 
together with detailed discussions), we mention that, upon Wick's 
rotation on the time integration ($\int_{-\infty}^\infty \to 
\int_{-i \infty}^{i\infty}$), our analytical solutions can be employed
explicitly. The final results for the above two induced condensates are
recorded immediately below:
$$\eqalignno{
 <0\mid {\bar q}(0) \gamma_\mu \gamma_5 q(0)\mid 0> & \equiv g \chi Z_\mu 
<{\bar q}q>, & (33a)\cr
 <0\mid {\bar q}(0) g_c {\tilde G}_{\mu\nu} \gamma^\nu q(0)\mid 0> 
&\equiv g \kappa Z_\mu <{\bar q}q>, &(33b)\cr 
\chi^{(1)} <{\bar q}q> &= {m^2\over 2 \pi^2} (\ln{\pi m^2\over \mu^2} 
   + \gamma_E), &(33c)\cr
\chi^{(2)} &= {\pi \over 2}{m\over \xi_0} - {1\over 16} {m m_0^2\over 
\xi_o^2} + {g_c^2\over 216} {<{\bar q}q>\over \xi_0^2}, &(33d)\cr
\kappa &= {m\over 8}(\pi + 2 \ln 2 +1) - {1\over 32} {m m_0^2 \over 
 \xi_0}. &(33e)}$$
Numerically, we find $\chi a \approx 0.15\, GeV^2$ and $ \kappa a\approx
6 \times 10^{-4} GeV^4$. (Here we have used the updated value on
the gluon condensate [7], the current quark mass, and other input 
parameters as adopted previously [4].) Such values for the induced 
condensates, albeit somewhat
smaller than the commonly adopted ones, are not unexpected since the 
leading contributions are linear in the ``current'' quark mass $m$. 

There are other problems for which our analytical results may be very
useful. For example, the amplitude given by
$$\eqalignno{T_{\mu\nu}(q^2,p\cdot q) &= i \int d^4x e^{-iq\cdot x} 
<\pi^+ (p)\mid T(J_\mu(x) J_\nu(0))\mid \pi^+(p)> &(34a)\cr
  &\equiv (-g_{\mu\nu} +{q_\mu q_\nu\over q^2})T_1 (q^2, p\cdot q)&\cr
  & \quad +{1\over p^2}(p_\mu -{p\cdot q\over q^2}q_\mu)(p_\nu -
  {p\cdot q\over q^2}q_\nu)T_2(q^2, p\cdot q), &(34b) }$$
characterizes the forward Compton scattering off $\pi^+$ (a Goldstone
boson) and also the parton distributions of $\pi^+$. Applying the 
soft-pion theorem (together with current algebra), we find, in the 
limit of $p_\mu \to 0$, 
$$\eqalignno{
 &T_{\mu\nu}(q^2,0)&\cr
=& {i\over f_\pi^2}\int d^4x e^{-iq\cdot x} <0\mid
 T(\{A_\mu^1(x)-i A_\mu^2(x)\}\{A_\nu^1(0)+iA_\nu(0)\}- 2 V_\mu^3(x)
    V_\nu^3(0))\mid 0>.&\cr
=& {i\over f_\pi^2}\int d^4x e^{-iq\cdot x} Tr\{ i S_d^{ba}(-x)\gamma_\mu
   \gamma_5 i S_u^{ab}(x)\gamma_\nu \gamma_5 &\cr
 &\qquad\qquad\qquad -{1\over 2}iS_u^{ba}(-x) \gamma_\mu iS_u^{ab}(x) 
  \gamma_\nu -{1\over 2} iS_d^{ba}(-x)\gamma_\mu iS_d^{ab}(x) \gamma_\nu\}.
   &(35)}$$
The structure functions $W_i(q^2, p.q)$ (in the description of deep 
inelastic scattering off the $\pi^+$ target) is the imaginary part of
$T_i(q^2, p\cdot q)$ divided by the factor $\pi$. Eq. (35) suggest that
our analytical expressions for the nonlocal condensates may be useful
for analyzing properties of Goldstone pions. We find, as a soft-pion limit,
$$\eqalignno{
 W_1(q^2, p\cdot q) &= {1\over \pi} \rm{Im}\, T_1(q^2, p\cdot q) &\cr
 & \longrightarrow {m<{\bar q}q>\over 2 f_\pi^2 \xi_0}, \quad
   \rm{as}\quad p_\mu\to 0\,\, \rm{and}\,\, q_\mu \to 0. &(36)}$$
This limit is derived making use of our analytical expressions on the
nonlocal condensates, again with Wick's rotation on the time integration.

In summary, we have in this paper suggested a specific way of obtaining
a closed set of coupled differential equations for nonlocal condensates.
Specifically, the leading-order equations for the nonlocal condensates 
in relation to the quark propagator are obtained and explicit analytical
solutions are obtained. We believe that such results could be very 
useful for a large number of problems in hadron physics.

\bigskip
\bigskip
\centerline{\bf Acknowledgement}
\medskip
\medskip

The author wishes to acknowledge Center for Theoretical Physics of MIT for
the hospitalities extended to him during his sabbatical leave. He also
acknowledges the National Science Council of R.O.C. for its partial support 
(NSC86-2112-M002-010Y) towards the present research.
 
\bigskip
\bigskip
 
\centerline{\bf References}

\item{1.} For notations, see, e.g., T.-P. Cheng and L.-F. Li, {\it Gauge 
Theory of Elementary Particle Physics} (Clarendon Press, Oxford, 1984).

\item{2.} M. A. Shifman, A.J. Vainshtein, and V.I. Zakharov, Nucl. Phys. 
{\bf 147}, 385, 448 (1979).

\item{3.} W-Y. P. Hwang, Preprint hep-ph/9601219 \& MIT-CTP-2498, Z. Physc.
{\bf C}, accepted for publication.

\item{4.} K.-C. Yang, W-Y. P. Hwang, E.M. Henley, and L.S. Kisslinger, 
Phys. Rev. {\bf D47}, 3001 (1993).

\item{5.} B. L. Ioffe, Nucl. Phys. {\bf B188}, 317 (1981); (E) {\bf B191}, 591
(1981); V. M. Belyaev and B. L. Ioffe, Zh. Eksp. Teor. Fiz. {\bf 83}, 876 
(1982) [Sov. Phys. JETP {\bf 56}, 493 (1982)].

\item{6.} V. M. Belyaev and Ya. I. Kogan, Pis'ma Zh. Eksp. Teor. Fiz. {\bf 37},
611 (1983) [JETP Lett. {\bf 37}, 730 (1983]; C. B. Chiu, J. Pasupathy, and S.J.
Wilson, Phys. Rev. {\bf D32}, 1786 (1985); E. M. Henley, W-Y. P. Hwang, and
L.S. Kisslinger, Phys. Rev. {\bf D46}, 431 (1992); Chinese J. Phys. (Taipei)
{\bf 30}, 529 (1992).

\item{7.} S. Narison, Phys. Lett. {\bf B} 387, 162 (1996).

\bye